\documentclass{ws-procs975x65}

\def\beq{\begin{equation}}
\def\eeq{\end{equation}}

\begin{document}

\title{Archimedes: a feasibility study of an experiment to weigh the electromagnetic vacuum}
\author{Enrico~Calloni$^2$, S.~Caprara$^3$, Martina~De~Laurentis$^2$, Giampiero~Esposito$^2$, M.~Grilli$^3$, 
E.~Majorana$^3$, G.~P.~Pepe$^2$, S.~Petrarca$^3$, Paola~Puppo$^*,^3$, P.~Rapagnani$^3$, Fulvio~Ricci$^3$, 
Luigi~Rosa$^2$, Carlo~Rovelli$^4$, P.~Ruggi$^1$, N.L.~Saini$^3$, Cosimo~Stornaiolo$^2$, F.~Tafuri$^2$}

\address{$^1$European Gravitational Observatory (EGO), Cascina (Pisa),\\$^2$University of Napoli Federico II and INFN Napoli,
\\$^3$University of Roma Sapienza and INFN Roma,\\$^4$University of Aix-Marseille,\\ $^*$E-mail: paola.puppo@roma1.infn.it }

\begin{abstract}
Archimedes is a feasibility study of a future experiment to ascertain the interaction of vacuum fluctuations 
with gravity. The experiment should measure the force that the earth’s gravitational field exerts on a Casimir 
cavity by using a small force detector. Here we analyse the main parameters of the experiment and we present 
its conceptual scheme, which overcomes in principle the most critical problems.
\end{abstract}


\bodymatter

\section{Introduction}
The cosmological constant problem is among the main fundamental unsolved questions of modern physics. 
The vacuum energy density resulting from quantum field theory is enormously larger than the value constrained from 
General Relativity, by considering the radius of our universe and its accelerated expansion. Although there is a long list 
of detailed and important theoretical works, e.g. \cite{Weinberg,DeWitt,Pad}, 
a real agreement has not been reached on this topic. 
The predicted effect is such that only in the last few years, the improved techniques for 
high temperature superconductors and small forces' detection 
have shown a way to follow a real experimental path \cite{Pos}.

\section{Weighing the vacuum energy}
The idea is to weigh the vacuum energy stored in a rigid Casimir cavity formed by parallel conducting plates. 
If the vacuum energy does interact with gravity, a force directed upwards acts on the cavity and can be interpreted 
as the lack of weight of the modes expelled by it, similarly to the Archimedes buoyancy of a fluid. 
The force is equal to \cite{Calloni}:
\begin{equation}
\vec{F}=-{\left| E_C\right|\over c^2}\vec{g}=-{\cal A}{\pi^2 \hbar\over 720~a^3}{\vec{g}\over c}
\end{equation}
where $E_C$ is the vacuum energy stored in the cavity, $a$ is the gap between the plates and $\cal{A}$ is their surface, 
$c$ is the speed of light and $\vec{g}$ is the earth's gravitational acceleration directed downwards.  

The measurement principle is to modulate the Casimir cavity weight by periodically changing its vacuum energy. 
This can be obtained by modulating the reflectivity of the plates. 
Using superconducting plates and moving around their transition temperature can be an effective 
method to change their reflectivity \cite{Calloni}. The use of cuprates like YBCO, a type-II high layered 
$T_c$ superconductor can be suitable for our purpose. The cuprates are formed by a huge number of natural stacks of 
superconductor layers separated by dielectric layers and they can be modeled as multi-Casimir cavities 
that provide a big  energy change during transition.
Even though a complete study is still lacking, and it is one of the problems to be faced inside the Archimedes 
project, the expected amplitude $F$ of the modulated force is of order $F\sim10^{-16} N$. 

\section{The experiment}
Either the gravitational wave detectors or the balances could be used for detecting the Archimedes force\cite{Calloni}. 
However, the detection bandwidth typical of the thermal processes in the periodical transition from normal to superconducting 
state is in the region of 1-100 mHz, so that the balances are more suitable for this application.

A very sensitive balance requires an excellent isolation from the seismic noise at the very low frequency regime 
of 1-100 mHz. The use of a seismic-attenuation device dealing with an inverted pendulum and a suitably tuned resonator 
hung to it can fullfil the requirements.
Another challenging point is the thermal modulation around the $T_c$ transition temperature. Only the radiative mechanism 
must be used to remove or add thermal energy to the sample, because 
it must be isolated from any external interaction that could add energy other than the vacuum one.
Thermal times depend on the thermal properties of chosen materials and their geometry, and since the radiative exchange 
is privileged, an optimization is needed to reduce the transition times and increase the frequency bandwidth. 
To follow all the requirements both the thermal and the mechanical design must proceed together. 

\section{Conclusion}
The Archimedes experiment aims at verifying the feasibility to measure the weigh of the vacuum. 
Three crucial points are focused with this project, the construction of a balance sensitive to forces 
at the level of $10^{-16} ~ N$, a thermal modulation system using only a radiative mechanism, and the detailed 
study of high $T_c$ superconducting systems as the main source of vacuum energy.

\end{document}